\documentclass[
 aip,apl,
 amsmath,amssymb,
 manuscript,
]{revtex4-1}

\usepackage{graphicx}
\usepackage{dcolumn}
\usepackage{bm}
\usepackage[mathlines]{lineno}

\usepackage[utf8]{inputenc}
\usepackage[T1]{fontenc}
\usepackage{mathptmx}

\begin{document}

\title[TACIT APL]{Demonstration of a Tunable Antenna-Coupled Intersubband Terahertz (TACIT) Mixer}

\author{C. Yoo}
 \affiliation{Physics Department and Institute for Terahertz Science and Technology, University of California, Santa Barbara, California 93106}
 
\author{M. Huang}
 \affiliation{Physics Department and Institute for Terahertz Science and Technology, University of California, Santa Barbara, California 93106}

\author{J. H. Kawamura}
 \affiliation{Jet Propulsion Laboratory, California Institute of Technology, Pasadena, California 91109}

\author{K. W. West}
 \affiliation{Department of Electrical Engineering, Princeton University, Princeton, New Jersey 08544}
 
\author{L. N. Pfeiffer}
 \affiliation{Department of Electrical Engineering, Princeton University, Princeton, New Jersey 08544}
 
\author{B. S. Karasik}
 \affiliation{Jet Propulsion Laboratory, California Institute of Technology, Pasadena, California 91109}
 
\author{M. S. Sherwin}
 \affiliation{Physics Department and Institute for Terahertz Science and Technology, University of California, Santa Barbara, California 93106}

\date{\today}

\begin{abstract}
A fast, voltage-tunable terahertz mixer based on the intersubband transition of a high-mobility 2-dimensional electron gas (2DEG) has been fabricated from a single 40 nm GaAs-AlGaAs square quantum well heterostructure. The device is called a Tunable Antenna-Coupled Intersubband Terahertz (TACIT) mixer, and shows tunability of the detection frequency from 2.52 THz to 3.11 THz with small (< 1 V) top gate and bottom gate voltage biases. Mixing at 2.52 THz has been observed at 60 K with a -3dB intermediate frequency (IF) bandwidth exceeding 6 GHz.

\end{abstract}

\maketitle

Terahertz (THz) heterodyne receivers are widely used for high-resolution THz spectroscopy in space and are important for applications in bio-medical imaging and future THz wireless communications.\cite{Siegel2002,Hubers2008} In a THz heterodyne system, a non-linear element referred as a mixer down-converts an incoming THz signal (RF) into an intermediate-frequency (IF) signal at a much lower frequency (usually at several GHz) using a THz local-oscillator (LO) source with known frequency and power. For a low-noise, high-resolution THz heterodyne system, a sensitive, broadband mixer is essential. Above 1 THz, superconducting hot-electron bolometers (HEBs) are current state-of-the-art mixers that offer low noise temperature (within an order of magnitude of the quantum limit), wide IF bandwidth ($\sim$3 GHz), and small required LO power ($\sim$1 $\mu$W).\cite{Hubers2008} However, cryogenic operating temperature less than or equal to 4 K limits the use of superconducting HEBs in certain applications such as deep-space missions to planets and comets that cannot afford the power and mass required for active cryogenic cooling. For such applications, Schottky diode mixers that work at ambient temperature are the only option so far, but with the cost of higher noise and higher required LO power ($\sim$1 mW).  The latter especially limits the use of Schottky mixers in heterodyne array applications.

In this Letter, we demonstrate a Tunable Antenna-Coupled Intersubband Terahertz (TACIT) mixer, which is a new type of THz mixer based on the intersubband transition of a high-mobility 2-dimensional electron gas (2DEG) confined in a quantum well.\cite{Sherwin1997} TACIT mixers operate at relatively high temperatures (50 - 70 K), which are accessible with passive cooling for deep-space missions and with compact, light-weight cryocoolers for other applications, and are predicted to offer low single-sideband (SSB) noise temperature ($\sim$1,000 K), a wide IF bandwidth (> 10 GHz), and low required LO power (< 1$\mu$W) along with tunability in the detection frequency (2-5 THz).\cite{Sherwin2002} The prototype TACIT mixer described in this Letter demonstrates tunability in the detection frequency between 2.52 THz and 3.11 THz with small top and bottom gate biases (< 1 V) and THz  mixing at 2.52 THz at 60 K with an IF bandwidth exceeding 6 GHz. The noise temperature, the conversion loss, and the required LO power of the prototype device have not been measured for the current non-optimized device. However, an analysis of the current-voltage (IV) curve for the active region of a model device indicates that a further optimized TACIT device will offer impressive mixer characteristics, making the TACIT mixer a viable, alternative mixer technology for low-noise, high-resolution THz spectroscopy with possible array applications in deep space, and for other applications in which relaxed cryogenic and LO power requirements are advantageous.

A TACIT mixer is a type of bolometric mixer based on the hot-electron effect of a high-mobility 2DEG. Since the early 1990s, two-terminal versions of 2DEG-based HEB mixers have been proposed,  \cite{Yngvesson2000} and have successfully demonstrated mixing in the millimeter-wave range with wide IF bandwidths ranging from 3 GHz for phonon-cooled devices\cite{Yang1995,Morozov2005} to 20 GHz for diffusion-cooled devices,\cite{Lee2001} and even to 40 GHz for ballistically-cooled devices.\cite{Lee2002} Like a superconducting HEB, a 2DEG HEB has an antenna coupled to two ohmic contacts (source and drain) that orients RF and LO electric fields along the 2DEG plane, and the radiation is absorbed through ohmic losses in the electrons. Because of the large kinetic inductance of high-mobility 2DEGs,\cite{Burke2000} however, the RF coupling efficiency in these two-terminal devices significantly degrades at THz frequencies. As a result, the conversion loss greatly increases above the practical upper frequency limit of 500 GHz,\cite{Lee2002} limiting the application of the two-terminal 2DEG HEB mixers at THz frequencies.

\begin{figure*}
\includegraphics[width=1\textwidth]{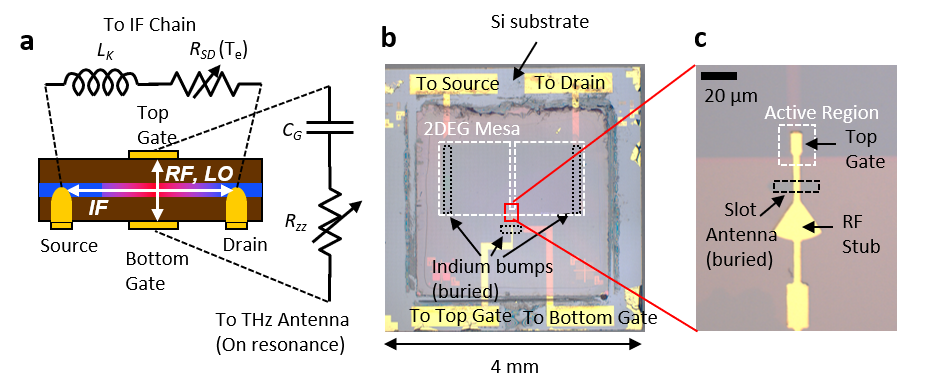}
\caption{\label{fig:fig1} (a) Schematic for the vertical profile of a TACIT mixer with equivalent circuits for the device impedance seen by the IF chain and by the THz antenna on resonance. A single GaAs-AlGaAs quantum well membrane (brown) contains a high-mobility 2DEG (color with gradation) with two ohmic contacts (source and drain) to measure IF response. Two Schottky gates (top and bottom gates) are used to couple THz radiation into the active region of the device. (b) Optical microscope image of a fabricated TACIT mixer showing the semi-transparent GaAs-AlGaAs quantum well membrane on a Si substrate that has Au bonding pads for electrical access to the four terminals. (c) Optical microscope image of the active region covered by the top gate metallization with an RF stub and choke filter.}
\end{figure*}

TACIT mixers overcome the frequency limit in two-terminal 2DEG HEB mixers by using two additional gates to achieve a high THz coupling efficiency through the intersubband transition of a 2DEG confined in a quantum well. In this four-terminal device scheme (see Figure \ref{fig:fig1}(a)), two ohmic contacts (source and drain) are used to apply a DC bias current and measure the IF response in the device resistance, and two Schottky gates (top gate and bottom gate) are used to couple THz RF and LO into the active region of the device in which THz radiation is resonantly absorbed by the 2DEG through an intersubband transition. The absorbed THz energy promotes the electrons from the ground subband to the first excited subband.  The electrons thermalize in less than 10 ps above 50K \cite{Heyman1995} and heat up the active region, resulting in a fast bolometric response in the device resistance that follows IF at GHz frequencies. To satisfy the selection rule for an intersubband transition in a confined 2DEG,\cite{Helm1999} a planar antenna structure is integrated in the bottom gate metallization to orient the THz electric fields perpendicular to the 2DEG plane.

With a top gate and a bottom gate to couple the THz radiation, the impedance of a TACIT mixer can be separately optimized for the THz antenna and for the IF chain. In two-terminal 2DEG HEB mixers, both the RF impedance seen by the antenna and the IF impedance for the IF chain are modelled by an inductor with the kinetic inductance of a 2DEG, $L_{k}$, in series with a resistor with the source-drain resistance $R_{SD}(T_{e})$ that depends on electron temperature $T_{e}$ (see the equivalent circuit for the IF chain in Figure \ref{fig:fig1}(a)). In TACIT mixers, however, the THz absorption occurs between the top gate and the bottom gate, and the RF impedance is now modelled by a capacitor with the geometric capacitance $C_{G}$, formed by the top and the bottom gate, in series with an effective resonator circuit that represents the intersubband transition.\cite{Sherwin2002} On resonance, the impedance of the effective resonator becomes purely resistive and is modelled by a resistor with the resistance $R_{zz}$ which is presented to a current oscillating between the top gate and the bottom gate. It is important to note that the resistance $R_{zz}$ is different from  the source-drain resistance $R_{SD}(T_{e})$, which is responsible for the IF response. The antenna structure is designed in a way that the real part of the antenna impedance roughly matches $R_{zz}$ and the reactive part tunes out $C_{G}$. In a fabricated device, $R_{zz}$ can be further tuned by varying the charge density with voltage biases on the top and bottom gates, and the RF impedance can be matched very closely to the antenna impedance, yielding a high RF coupling efficiency in a TACIT mixer.\footnote{See Supplementary Material for further details on the impedance matching in TACIT mixers.} 

 The top gate and the bottom gate are also used to tune the intersubband absorption frequency that determines the detection frequency in a TACIT mixer. Voltage-tunable direct detection via intersubband absorption has already been demonstrated in various 2DEG systems including a Si inversion layer\cite{Wheeler1971} and coupled GaAs-AlGaAs quantum well structures.\cite{Tomlinson2000,Serapiglia2005} In the TACIT mixer demonstrated here, the detection frequency is determined by the absorption frequency for the intersubband transition between the ground and first excited subbands of a 2DEG confined in a single, square GaAs-AlGaAs quantum well that is 40 nm wide.  A wide tunability in the absorption frequency (2-5 THz) of such a well has been observed previously.\cite{Williams2001} The absorption frequency, which is primarily determined by the width of the square quantum well, further depends on the charge density in the active region $n_{s}$ and the DC electric field $E$ applied perpendicularly to the 2DEG (along the growth direction of the quantum well).\cite{Williams2001} These two quantities are independently controlled by biasing the top gate and the bottom gate using the following relations\footnote{These relations assume that the two metal gates form parallel-plate capacitors with the 2DEG.}
\begin{equation}\label{eq1}
n_{s}=n_{0}+\frac{c}{e}(V_{T}+V_{B})    
\end{equation}
and
\begin{equation}\label{eq2}
E=\frac{1}{d}(V_{T}-V_{B})   
\end{equation}
where $n_{0}$ is the intrinsic charge density in the active region without any gate biasing, $c$ is the capacitance per unit area between each gate and the 2DEG, $e$ is the elementary charge, $V_{T(B)}$ is the top (back) gate voltage, and $d$ is the distance between each gate and the 2DEG.

\begin{figure}
\includegraphics[width=0.5\textwidth]{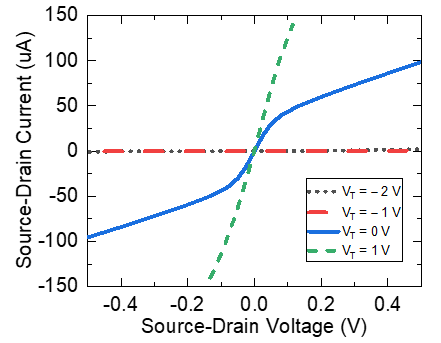}
\caption{\label{fig:fig2} Current-voltage (IV) curves of the fabricated TACIT mixer at 50 K with different charge densities in the active region, showing the tunable bolometric response. The bottom gate was fixed at 0 V. At V\textsubscript{T} = -2 V and V\textsubscript{T} = -1 V (dotted and long-dashed lines), no current flows as the active region is depleted of electrons. At V\textsubscript{T} = 0 V (solid line), there are enough electrons such that the current becomes non-linear at a higher bias when the electrons near the active region start to heat up. At V\textsubscript{T} = 1 V (short-dashed line), there are excess electrons that lead to the linear behavior in the current in the given bias range.}
\end{figure}

A prototype TACIT mixer was fabricated from a modulation-doped 40 nm GaAs-AlGaAs square quantum well heterostructure (see Figures \ref{fig:fig1} (b) and (c)). The sample was grown by Molecular Beam Epitaxy (MBE) and contains a high-mobility 2DEG with mobility $\mu= 9.4\times10^{6}$ cm\textsuperscript{2} V\textsuperscript{-1} s\textsuperscript{-1} and charge density $n_{0}=2.2\times 10^{11}$ cm\textsuperscript{-2} at 2 K.\footnote{For more details on sample structure and device fabrication, see Supplementary Material} To process both sides of the quantum well sample, a modified version of the EBASE flip-chip process\cite{WECKWERTH1996561} with UV contact lithography was used.\footnote{For more details on the fabrication, see Supplementary Material.} For the prototype device, a single-slot antenna (see Figure \ref{fig:fig1} (c)) was chosen for its simplicity in design and capability of providing sufficient THz coupling to linearly polarized sources despite the asymmetric beam pattern in horizontal and vertical polarizations. 

After the fabrication, current-voltage (IV) curves were measured to check the bolometric nature of the detection mechanism and the gating of the fabricated device, and both direct detection and heterodyne detection were performed to verify the tunability in the THz response and the capability of THz mixing, respectively. For each IV curve measurement, the source was biased with the drain being grounded, and the top gate and the bottom gate were biased to a fixed voltage to avoid floating gates that would otherwise make the charge density in the active region fluctuate during the measurement. For direct and heterodyne detection measurements, a hyperhemispherical Si lens was used to quasi-optically couple THz radiation into the active region of the device. The device was bonded on the back side of the Si lens, and the mixer block that contains the Si lens was thermally anchored to a cold plate in a liquid helium cryostat which could be warmed up to 60 K. For direct detection, a CO\textsubscript{2} pumped molecular gas far-infrared (FIR) laser provided THz signals at 2.52 THz and 3.11 THz. For heterodyne detection, THz RF was provided by a custom tunable solid-state frequency multiplier built at Jet Propulsion Laboratory, and THz LO was provided by the FIR laser. A bias-tee was used to apply DC bias to the source-drain channel and to couple out the IF response, which was amplified with a low-noise amplifier and detected by a lock-in amplifier. For both THz detection measurements, the source was biased to 50 mV with the drain being grounded.

\begin{figure}
\includegraphics[width=0.5\textwidth]{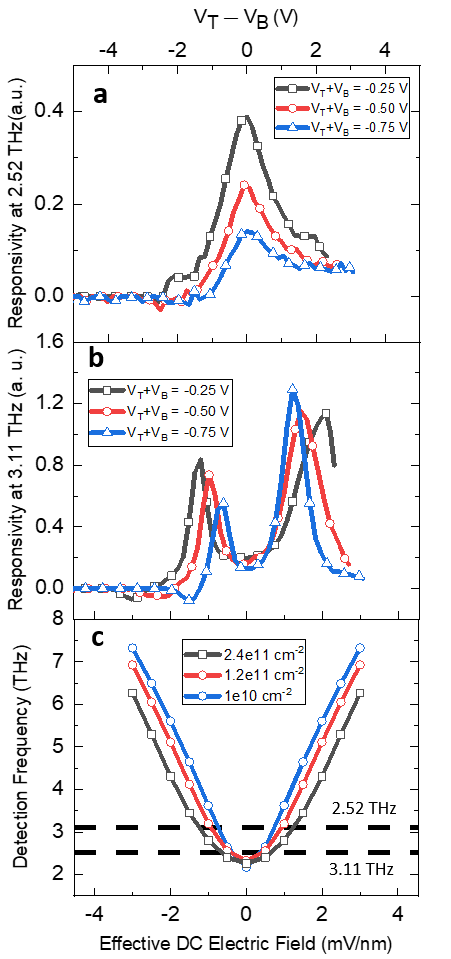}
\caption{\label{fig:fig3} (a) Responsivity at 2.52 THz at 36 K as a function of the effective DC electric field applied to the active region along the growth direction of the quantum well. Each curve corresponds to a different charge density set by voltage sum ($V_{T}+V_{B}$). (b) Responsivity at 3.11 THz at 36 K as a function of the effective DC electric field. The responsivity behavior beyond the DC electric field value of 1 mV/nm is due to the breakdown of the top gate above its threshold voltage. (c) Calculated detection frequency of the fabricated TACIT mixer at 36 K as a function of the  DC electric field. Each curve corresponds to an estimated charge density in the active region.}
\end{figure}

Figure \ref{fig:fig2} shows the IV curves of the TACIT mixer at 50 K with different charge densities in the active region. The top gate voltage was varied from -2 V to 1 V with the bottom gate fixed to 0 V to effectively vary the charge density, which is set by the voltage sum ($V_{T}+V_{B}$; see Eq. \ref{eq1}). We observed that the behavior of the current flow in the device was tunable from being completely flat (when the active region is depleted of electrons) to being linear (when there are excess electrons). At V\textsubscript{T} = 0 V, the IV curve showed a non-linear bolometric response due to the hot-electron effect. The temperature coefficient of resistance  $\alpha$ ($\alpha = \frac{1}{R}\frac{dR}{dT}$) was measured to be $\sim 0.02$ K\textsuperscript{-1}.\footnote{For further details on the temperature dependence of the bolometric response, see Supplementary Material.} We observed similar tunability in the IV curve with voltage biases on the bottom gate, confirming the normal operation of both gates within the tested range of the voltage biases. The device resistance at the source-drain bias of 50 mV was $\sim$500 $\Omega$ at V\textsubscript{T} = 1 V. Compared with the 2DEG sheet resistivity of 75 $\Omega/\Box$ measured at 50 K in a Hall bar sample, the high device resistance at V\textsubscript{T} = 1 V was expected due to the non-square 2DEG mesa geometry and the corresponding current-crowding effect that occurs when current flows from the 1 mm wide 2DEG connected to the contacts to the 5 $\mu$m wide active region of the device. With optimal mesa design and proper tuning of the charge density, we expect that the device resistance can be lowered to better match the 50 $\Omega$ impedance of the IF chain.

Figure 3 shows the direct detection results at 2.52 THz and 3.11 THz at 36 K (Figures \ref{fig:fig3} (a) and (b)) along with the calculated detection frequency at the same temperature (Figure \ref{fig:fig3} (c)). Figures \ref{fig:fig3}(a) and \ref{fig:fig3}(b) show the device responsivities at 2.52 THz and 3.11 THz as a function of the effective DC electric field, which was converted from the voltage difference ($V_{T}-V_{B}$; marked on the top of the plot) using Eq. 2. The curves were shifted horizontally with an offset of -0.98 mV/nm to cancel out a built-in electric field in the quantum well. Each curve in the plot corresponds to a different charge density in the active region set by the voltage sum ($V_{T}+V_{B}$). The voltage sum was not converted to a charge density as the charge density estimated from Eq. 1 was too high, possibly because of an asymmetry in the gate structure or the diffusion of the electrons out of the active region, and this discrepancy will be further investigated in the future. The theoretical model for the detection frequency shown in Figure \ref{fig:fig3}(c) is based on the calculation of the absorption frequency for the transition between the ground and the first excited subbands of a 2DEG confined in a 40 nm GaAs-AlGaAs square quantum well. The calculation includes the effect of the electrostatic potential from the charge density (self-consistent Hartree potential), many-body effects on the energy of the 2DEG (the exchange and correlation energies), and collective effects on the absorption (depolarization and exciton shifts).\cite{Helm1999}

The tunability in the detection frequency has been observed at 2.52 THz and at 3.11 THz (Figures \ref{fig:fig3}(a) and \ref{fig:fig3}(b)) with the absorption behavior consistent with the theoretical model (Figure \ref{fig:fig3}(c)). The range of experimentally accessible THz frequencies was limited by the available lines in the CO\textsubscript{2}-pumped molecular gas laser, and the device response beyond 3.11 THz will be investigated in the future with other THz sources such as a quantum cascade laser (QCL) to verify the tunable range for the detection frequency expected in the current design of the quantum well (2.5-4 THz). At 2.52 THz, we observed a single peak in the responsivity with the width of the peak increasing with a higher charge density. This behavior is consistent with the theoretical model shown in Figure \ref{fig:fig3}(c); the bottom dashed line at 2.52 THz intersects the bottom of the curves over a range of the DC electric field values with the range getting broader with increasing charge density. At 3.11 THz, we observed a double-peak behavior with the position of the first peak shifting to a more negative DC field value with increasing charge density. The shift in the peak position is also consistent with the model as the top dashed line at 3.11 THz (see Figure \ref{fig:fig3}(c)) intersects the curve for a higher charge density at a more negative DC electric field value. Above 1 mV/nm, the top gate starts to leak because it has exceeded its threshold voltage of $\sim$1 V. This leakage likely causes the asymmetry in the two peaks at 3.11 THz, and the non-zero responsivity above 1 mV/nm at 2.52 THz. The gate leaks because the top gate metal pad was deposited directly on the semi-insulating GaAs cap layer and forms a Schottky gate. In future devices, we can avoid the breakdown by having a thin oxide layer between the gate metal and the GaAs cap layer.

\begin{figure}
\includegraphics[width=0.5\textwidth]{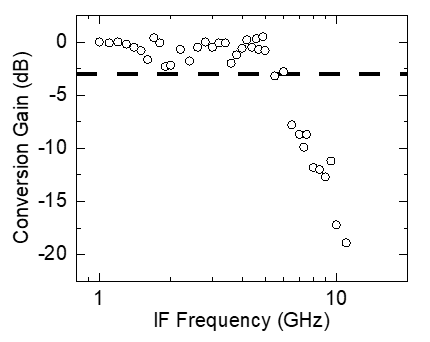}
\caption{\label{fig:fig4} Normalized IF response  at 2.52 THz at 60 K as a function of the IF frequency. The dashed line indicates the -3dB point.}
\end{figure}

Figure \ref{fig:fig4} shows the result of the heterodyne detection at 2.52 THz at 60 K. We observed IF signals in the GHz range, confirming the THz mixing capability of TACIT mixers. The frequency dependence of the IF response shows that the -3dB bandwidth exceeds 6 GHz. The frequency spectrum was not fitted with the single-pole Lorentzian because of the 12dB/octave roll-off that suggests the existence of higher-order filtering in the IF circuit. The -3dB bandwidth exceeding 6 GHz suggests that the fabricated TACIT mixer might be a diffusion-cooled device rather than a phonon-cooled device, in which the typical value of the -3dB bandwidth is $\sim$3 GHz in GaAs-AlGaAs 2DEG devices.\cite{Yang1995,Morozov2005} This is also consistent with the fact that the length of the active region of the TACIT mixer is 5 $\mu$m at which a diffusion-cooled device based on a high-mobility 2DEG was demonstrated previously.\cite{Lee2001}

Lastly, we estimate the noise temperature, the conversion loss, and the required LO power for an optimized TACIT mixer at 60 K by applying bolometric mixer theory\cite{Arams1966, Karasik,Karasik1996} to the IV curve for the active region of a model device. The IV curve for the active region was scaled from the IV curve measured in a Hall bar sample to make the device resistance $\sim$50 $\Omega$ at the bias point.\footnote{For more details on the scaling, see Supplementary Material.} For the modeling, we assumed the optimal bias, perfectly matched IF channel, and the RF coupling efficiency of 70\%, which is a typical value for hot-electron bolometers coupled with planar antenna structures (see, for example, Ref. \onlinecite{BorisCoupling}). At 60 K, the conversion loss is estimated to be $\sim$7.5 dB at a DC bias of $\sim$2.5 mV at which the dissipated DC power is $\sim$0.15 $\mu$W. For the noise temperature, we consider the contributions from Johnson noise and from thermal energy fluctuation. At the LO power of $\sim$0.2 $\mu$W, the Johnson mixer noise temperature and the thermal fluctuation mixer noise temperature are estimated to be $\sim$400 K and $\sim$300 K respectively, yielding the total single-sideband (SSB) mixer noise temperature of $\sim$700 K.\footnote{For more details on the mixer modeling, see Supplementary Material.}

In summary, we have achieved a new type of THz heterodyne mixer based on the intersubband transition of a high-mobility 2DEG confined in a 40 nm GaAs-AlGaAs square quantum well. The fabricated device shows THz response at 2.52 THz and 3.11 THz consistent with theoretical predictions, demonstrating the tunability of TACIT mixers. Heterodyne detection at 2.52 THz was observed at 60 K with an IF bandwidth exceeding 6 GHz, showing the THz mixing potential of TACIT mixers with a wide IF bandwidth. 

\begin{acknowledgments}
This research was carried out in part at the Jet Propulsion Laboratory, California Institute of Technology, under a contract with the National Aeronautics and Space Administration. A portion of this work was performed in the UCSB Nanofabrication Facility, an open access laboratory. We thank Karl Unterrainer and Michael Krall for early help with ohmic contacts, and Dr. Brian Thibeault and Dr. Demis D. John for their help in discussing the flip-chip processing steps. The C++ codes for calculating the detection frequency of TACIT mixers were written by Dr. Bryan Galdrikian and modified by Dr. Chris Morris. We are grateful for early leadership by W. R. McGrath (deceased).  The work at UCSB was supported by NASA PICASSO program via a contract with JPL. 

\end{acknowledgments}

\section*{supplementary material}

\section{RF Impedance Matching}
In TACIT mixers, the RF impedance of the active region seen by the THz antenna can be calculated by considering a current oscillating between the top gate and the bottom gate and the corresponding voltage drop caused by the polarization of the 2DEG\cite{Sherwin2002}. The expression for the RF impedance $Z(\omega)$ is given in Ref. \onlinecite{Sherwin2002}: 
\begin{equation}\label{seq1}
    Z(\omega) = \frac{1}{i\omega\epsilon\epsilon_{0}A}\left(d-\frac{\chi_{2DEG}(\omega)}{\epsilon\epsilon_{0}}\right)
\end{equation}
where $\omega$ is the angular frequency of incoming radiation, $\epsilon$ is the dielectric constant, $\epsilon_{0}$ is the permittivity of free space, $A$ is the area of the metal gates, $d$ is the distance between the top gate and the bottom gate, and $\chi_{2DEG}(\omega)$ is the 2D susceptibility of the 2DEG associated with intersubband transitions in quantum wells.

\begin{figure}[h]
\includegraphics[width=0.5\textwidth]{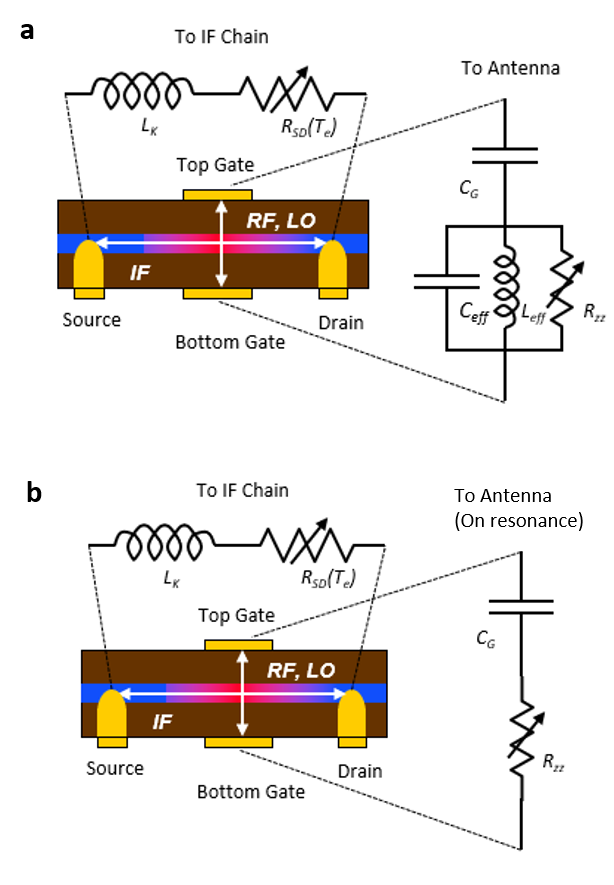}
\caption{\label{fig:sfig1to2}(a) Schematic for the vertical profile of a TACIT mixer with equivalent circuits for the IF impedance seen by the IF chain and the RF impedance seen by the THz antenna. The IF impedance is modelled by an inductor with the kinetic inductance of the 2DEG, $L_{k}$, in series with a resistor with the source-drain resistance $R_{SD}(T_{e})$ that depends on the electron temperature $T_{e}$. The RF impedance is modelled by a capacitor with the geometric capacitance $C_{G}$ ( formed by the top gate and the bottom gate) in series with an effective resonator circuit associated with the intersubband transition. (b) On resonance, the impedance of the resonator circuit becomes purely resistive, and the RF impedance can be modelled by a capacitor in series with a resistor with the resistance $R_{zz}$.}
\end{figure}

For TACIT mixers, we consider the 2D susceptibility $\chi_{2DEG}(\omega)$ associated with the intersubband transition between the ground and first excited subbands. The theory for $\chi_{2DEG}(\omega)$\cite{Allen1976,Ando1982} takes into account the effects of the dynamic screening and the electron-electron interaction through two simple parameters, the absorption frequency $\omega^*$ for the intersubband transition and the oscillator strength $f_{12}=2m^*z_{12}^2E_{12}/\hbar^2$ where $m^*$ is the effective mass of an electron in GaAs, $z_{12}$ is the matrix element of the position operator applied on the wavefunctions for subbands 1 and 2, $E_{12}$ is the energy spacing between the ground state and the first excited state, and $\hbar$ is the reduced Planck's constant.\cite{Sherwin2002} The expression for $\chi_{2DEG}(\omega)$ is given in Ref. \onlinecite{Sherwin2002} and Ref. \onlinecite{Ando1982}: 
\begin{equation}
    \label{seq2}
    \chi_{2DEG}(\omega) = \frac{N_{s}e^{2}f_{12}n(T_{e})}{m^{*}}\frac{1}{{\omega^{*}}^{2}-\omega^{2}+i2\omega\Gamma}
\end{equation}
where $N_{s}$ is the total sheet charge density, $e$ is the electron charge, $n(T_{e})$ is the normalized population difference $n(T_{e})=\frac{N_{1}(T_{e})-N_{2}(T_{e})}{N_{s}}$ where $N_{1(2)}(T_{e})$ is the population of subband 1 (subband 2) at the electron temperature $T_{e}$, and $\Gamma/2\pi$ is the HWHM of the intersubband absorption. With this expression for $\chi_{2DEG}(\omega)$, the device impedance $Z(\omega)$ can be modelled by an equivalent circuit of a capacitor with the geometric capacitance $C_{G}=\frac{\epsilon\epsilon_{0}A}{d}$ (formed by the two gates) in series with an effective resonator circuit that represents the intersubband absorption (Figure \ref{fig:sfig1to2}(a)). The resonator circuit consists of an inductive element with the effective inductance $L_{eff}$, a capacitive element with the effective capacitance $C_{eff}$, and a resistor with the resistance $R_{zz}$ presented to the current oscillating between the top gate and the bottom gate. It is important to note that the resistance $R_{zz}$ is different from the source-drain resistance $R_{SD}(T_{e})$, which is responsible for the IF response in a TACIT mixer.

When the frequency of the incoming radiation $\omega$ matches the absorption frequency for the intersubband absorption $\omega^{*}$ ($\omega=\omega^*)$, the reactive elements, $C_{eff}$ and $L_{eff}$, in the effective resonator circuit (see Figure \ref{fig:sfig1to2}(a)) tune each other out, and the resonator impedance become purely resistive (see Figure \ref{fig:sfig1to2}(b)). Hence, on resonance, the RF impedance $Z(\omega=\omega^*)$ can be modelled by a capacitor with the geometric capacitance $C_{G}$ in series with a resistor with the resistance $R_{zz}$:
\begin{equation}
    \label{seq3}
    Z(\omega^*)=\frac{d}{i\omega^*\epsilon\epsilon_{0}A}+\frac{N_{s}e^{2}f_{12}n(T_{e})}{\epsilon^{2}\epsilon_{0}^{2}A m^{*}{\omega^{*}}^{2} 2\Gamma} = \frac{1}{i\omega^*C_{G}}+R_{zz}
\end{equation}
where $R_{zz}$ is the effective resistance to the current oscillating between the top gate and the bottom gate. In a real device, the resistance $R_{zz}$ is tunable as the charge density in the active region can be varied by applying voltage biases to the top and bottom gates. With a combination of optimized design parameters (the gate metal area $A$ and the gate distance $d$) and the tunable $R_{zz}$, the RF impedance of a TACIT mixer can be matched very closely to the antenna impedance for a high coupling efficiency. 

\begin{figure*}
\includegraphics[width=0.9\textwidth]{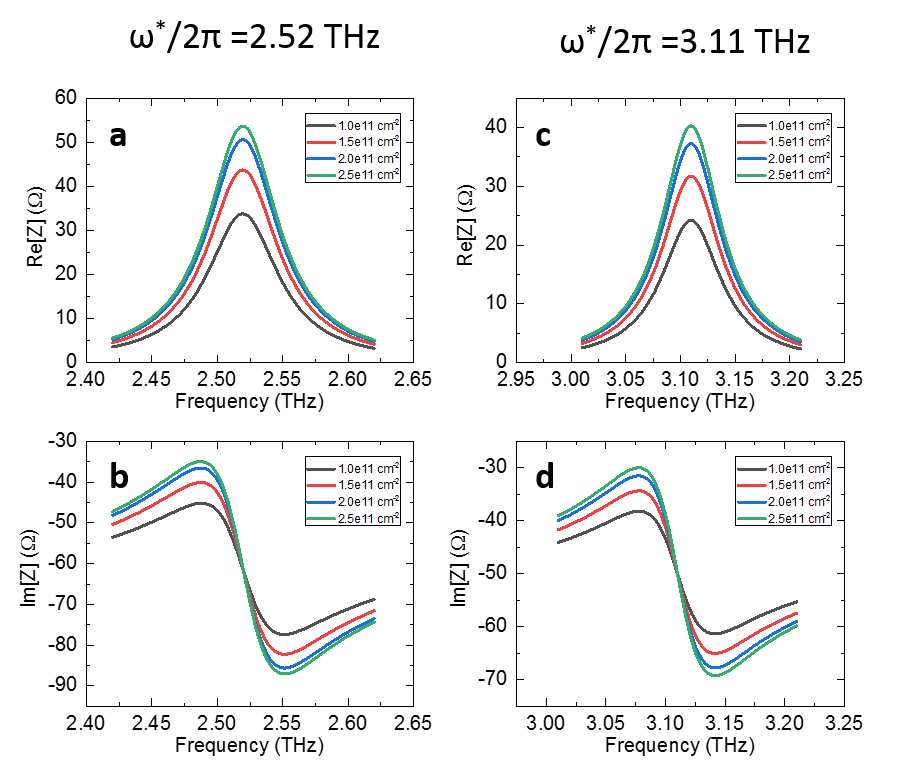}
\caption{\label{fig:Z_model_optimal} Calculated real part and imaginary part for the RF impedance for a model device with the gate area $A=$ 4 $\mu$m\textsuperscript{2} and the gate distance $d=$ 450 nm near the absorption frequency of 2.52 THz (subplots (a) and (b)) and near 3.11 THz (subplots (c) and (d)). For the calculation, the effective oscillator strength $f_{12}n(T_{e})$ was calculated for each charge density at the electron temperature of 85 K, which is the estimated temperature at the optimal bias point at the lattice temperature of 60 K. $\Gamma/2\pi=$ 33 GHz was used. With the optimization of the design parameters $A$ and $d$, the resistive part of the RF impedance can be designed to be near 50-70 $\Omega$, which are the typical values for the resistive impedance of planar antenna structures on a Si substrate. The imaginary part can be tuned off with an inductive element in the antenna circuit. }
\end{figure*}

With Eq. \ref{seq3}, we can estimate the RF impedance for the current prototype device (with the gate area $A=$ 20 $\mu$m\textsuperscript{2} and the gate distance $d=$ 660 nm) in the range of the charge density practically achievable in the device ($0.5\times10^{11}$ cm\textsuperscript{-2} $< N_{s} <2.5 \times10^{11}$ cm\textsuperscript{-2}). For the calculation, the effective oscillator strength $f_{12}n(T_{e})$ was calculated for each charge density, and the HWHM for the intersubband transition was assumed to be 33 GHz ($\Gamma/2\pi =$ 33 GHz) based on Ref. \onlinecite{Sherwin2002}. On resonance at 2.52 THz, the real part of the RF impedance for the prototype device is estimated to be 6-11 $\Omega$ in the given range for the charge density, and the imaginary part is estimated to be $-18 \Omega$. At 3.11 THz, the real part is 5-8 $\Omega$ and the imaginary part is  $-15 \Omega$. The non-zero values for the imaginary part are due to the geometric capacitance and can be tuned out with an inductive element in the antenna circuit in a future device. While the values for $R_{zz}$ in the prototype device are much smaller than typical values (50-70 ohm) for the resistive impedance of a planar antenna on a Si substrate, further optimization in the design parameters $A$ and $d$ can make the $R_{zz}$ to be close to 50-70 $\Omega$ (see Figure \ref{fig:Z_model_optimal}). This also leads to an increase in the imaginary part of the RF impedance on resonance, which can be tuned out with an inductive element in the antenna circuit.

\begin{figure*}
\includegraphics[width=0.9\textwidth]{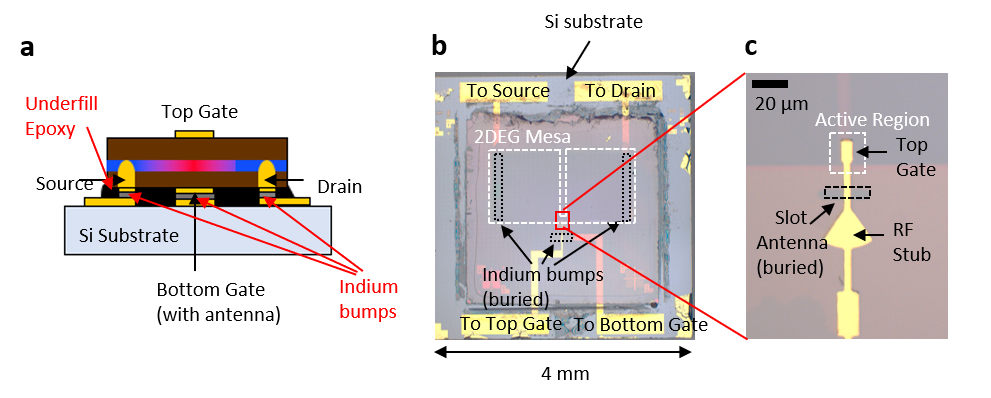}
\caption{\label{fig:sfig3} (a) Schematic for the vertical profile of a TACIT mixer. After the fabrication of the two ohmic contacts (source and drain) and the bottom gate, the quantum well sample is flip-chip bonded to a Si substrate with indium bump bonds. The Si substrate contains Au bonding pads that are electrically connected to the source, the drain, the bottom gate, and the top gate. (b) The optical microscope image of the fabricated TACIT mixer. (c) Optical microscope image showing the active region of the device.}
\end{figure*}

\section{Sample Structure and Fabrication}
The prototype TACIT mixer was fabricated from a modulation-doped, single 40 nm GaAs-AlGaAs square quantum well structure. The width of the quantum well, as well as the doping density, were set to make the detection frequency of the device to be near 2.5 THz at 60 K in the absence of additional charge density and external DC electric field. The quantum well sample was grown by Molecular Beam Epitaxy on a 2"-diameter, $\sim$500 $\mu$m thick GaAs substrate wafer.  A 500 nm GaAs buffer layer and a smoothing layer consisting of 100 periods of 10 nm Al\textsubscript{0.3}Ga\textsubscript{0.7}As and 3 nm GaAs were first grown on the substrate. For double-sided processing, 500 nm of Al\textsubscript{0.74}Ga\textsubscript{0.26}As was grown as an etch-stop layer. The growth for the single 40 nm quantum well followed, starting with a 10 nm GaAs cap layer, and the first barrier layer consisting of 200 nm Al\textsubscript{0.3}Ga\textsubscript{0.7}As, a Si modulation doping layer, and 100 nm Al\textsubscript{0.3}Ga\textsubscript{0.7}. Then, 40 nm GaAs quantum well layer was grown, followed by the second barrier layer consisting of a 100 nm Al\textsubscript{0.3}Ga\textsubscript{0.7}As layer, a Si modulation doping layer, and a 200 nm Al\textsubscript{0.3}Ga\textsubscript{0.7}As layer. Finally, another 10 nm GaAs cap layer was grown on the top surface. This created a single 40 nm quantum well with the center of the 2DEG layer 330 nm below the top surface. Hall measurements showed that the 2DEG in the sample has intrinsic charge density $n_{0}=2.2\times 10^{11}$ cm\textsuperscript{-2} and mobility $\mu=2.0\times10^{5}$ cm\textsuperscript{2} V\textsuperscript{-1} s\textsuperscript{-1} at 77 K, and $n_{0}=2.1\times 10^{11}$ cm\textsuperscript{-2} and $\mu=9.4\times10^{6}$ cm\textsuperscript{2} V\textsuperscript{-1} s\textsuperscript{-1} at 2 K.

We used a modified version of the EBASE flip-chip process\cite{WECKWERTH1996561} with UV contact lithography to fabricate both sides of the sub-micron thick quantum well structure (Figure \ref{fig:sfig3}(a)). To fabricate the front-side (later to be the buried side after flip-chip bonding), we first created $\sim$400 nm deep 2DEG mesa that consists of two squares with width $W=1$ mm and length $L=1$ mm connected by a narrow active region with $W=5$ $\mu$m and $L=4$ $\mu$m (see Figure \ref{fig:sfig3}(b)). Two 1 mm wide ohmic contacts were formed by evaporating a metal sequence of Ni-AuGe-Ni-Au and rapidly annealing the sample at 500 $^{\circ}$C in forming gas for 5 minutes. The specific contact resistance of the ohmic contact was measured in a transfer length method (TLM) pattern to be 3 $\Omega\cdot$mm in low temperatures (10 K$-$100 K). A bottom gate metal pad ($W=7$ $\mu$m and $L=4$ $\mu$m) was formed by evaporating Ti-Au to cover the active region, and the whole surface was passivated with a SiO\textsubscript{2} dielectric layer. For the deposition of indium, openings for the two ohmic contacts and the bottom gate pad were made by dry etching. A metal lead pattern that contains a planar single-slot antenna was deposited by evaporating Ti-Au on the SiO\textsubscript{2} layer to electrically contact the bottom gate metal pad. The single-slot antenna was chosen for design simplicity and can be replaced with a different antenna design for a better radiation pattern. 

For the back-side fabrication, we flip-chip bonded the processed quantum well sample to a Si substrate that has matching Au patterns with bonding pads (see Figure \ref{fig:sfig3}(b)). For the bonding, $\sim$1 $\mu$m thick indium bumps were thermally evaporated on the two ohmic contacts and on the lead pattern (Figure \ref{fig:sfig3}(b)). A Finetech flip-chip bonder was used to align the bonding surfaces of the quantum well sample and the Si substrate, and the sample and the substrate were pressed against each other while being heated up to 200$^{\circ}$C for 1 min. This made the indium bumps to reflow and bond with the matching Au patterns on the Si substrate, directly establishing the electrical connection between the three terminals on the buried side of the quantum well sample and the Au bonding pads on the Si substrate. EPO-TEK 353ND underfill epoxy was dispensed to fill the gap between the bonded surfaces and cured to protect the surfaces throughout the rest of the fabrication steps. For the top gate, the bulk ($\sim$450 $\mu$m) of the exposed GaAs substrate was lapped mechanically and the remaining GaAs layer ($\sim$50 $\mu$m) was removed with selective wet etch using citric acid until the etch was stopped by the Al\textsubscript{0.74}Ga\textsubscript{0.26}As etch-stop layer. This Al\textsubscript{0.74}Ga\textsubscript{0.26}As etch-stop layer was in turn removed by another selective wet etch using BHF until the etch was stopped by the 10 nm GaAs cap layer. This left the unprocessed side of the GaAs-AlGaAs membrane ready for the last contact lithography step in which the pattern for the top gate and the RF choke filter with the RF stub was defined (see Figure \ref{fig:sfig3}(c)). After the lithography, Ti-Au was evaporated to define the top gate metallization. Further details on the processing steps and the characterization of the mobility and the charge density after the flip-chip bonding and the back-side removal will be discussed in a future publication.

\section{Temperature Dependence of Bolometric Response}
The bolometric response of the prototype TACIT mixer (Device A) described in the main text was further investigated in another TACIT device (Device B) at 30 K - 100 K. Device B has the same design as Device A and was fabricated from the same 40 nm GaAs-AlGaAs quantum well sample, but was processed in a separate fabrication run.

\begin{figure*}
    \includegraphics[width=1\textwidth]{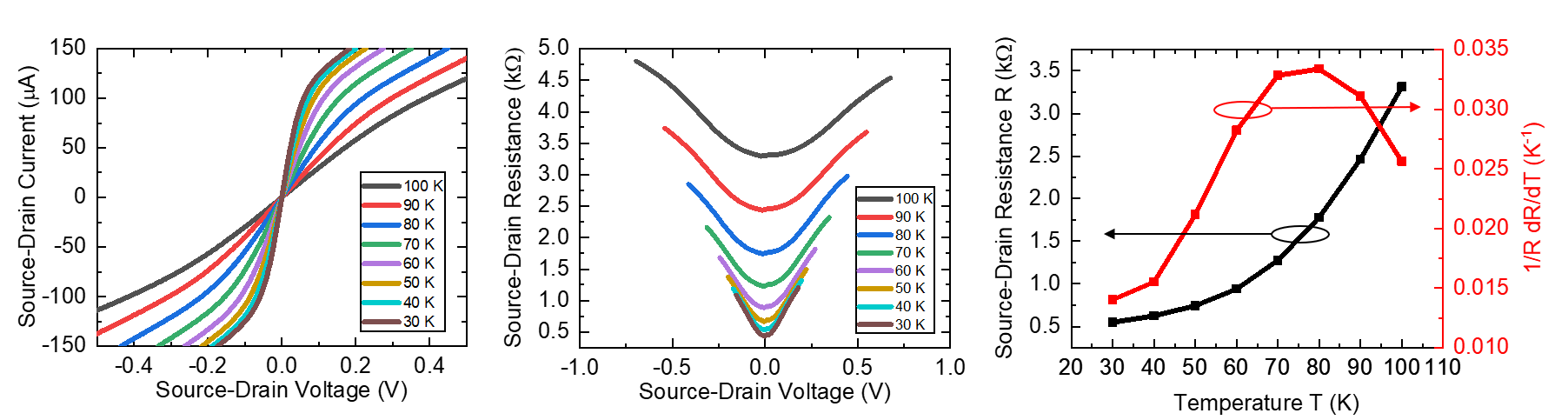}
    \caption{(a) Current-voltage (IV) curves for Device B at 30 K - 100 K. The top and the bottom gates were biased to 0.3 V. (b) Source-drain resistance of Device B as a function of the source-drain voltage at 30 K - 100 K. The top and the bottom gate were biased to 0.3 V. (c) Source-drain resistance (black curve) and the temperature coefficient of resistance (red curve) at 50 mV source-drain bias as a function of the  lattice temperature T.}
    \label{fig:sfig4to6}
\end{figure*}

Figures \ref{fig:sfig4to6}(a) and \ref{fig:sfig4to6}(b) show the IV curves and the resistance-voltage curves of Device B taken at 30 K - 100 K with $V_{T}+V_{B}$=0.3 V. Figure \ref{fig:sfig4to6}(c) shows the resistance (black curve) and the temperature coefficient of resistance $\alpha = \frac{1}{R}\frac{dR}{dT}$(red curve) at 50 mV source-drain bias as a function of the lattice temperature T. The high source-drain resistance (> 500$\Omega$) compared with the typical IF impedance of  the IF chain (50 $\Omega$) is due to the non-square geometry of the prototype TACIT device and the corresponding current-crowding effect caused by the current flowing from 1 mm wide contacts into the 5 $\mu$m active region. In a future device, the 2DEG mesa geometry can be optimized to minimize the current-crowding effect and the charge density to make the source-drain resistance to be near 50 $\Omega$ to match the impednace of the IF chain better.

\section{Mixer Modelling}

\begin{figure*}
    \includegraphics[width=0.8\textwidth]{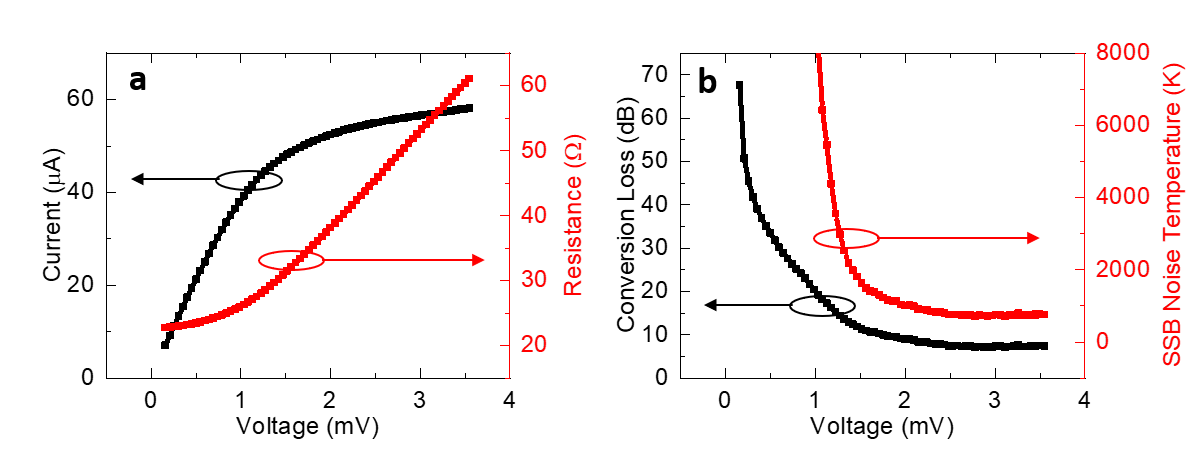}
    \caption{(a) The scaled IV curve for the active region of a model TACIT device at 60 K. The IV curve was scaled from the IV curve taken at 60 K to make the device resistance $\sim$50 $\Omega$ at the bias point. (b) Conversion loss (black) and SSB mixer noise temperature (red) extracted from the model IV curve using the bolometric mixer theory.}
    \label{fig:sfig7to8}
\end{figure*}

Mixer characteristics such as the noise temperature, the conversion loss, and the optimal LO power were estimated from a model IV curve for the active region of a TACIT mixer at the lattice temperature of 60 K. At the bias point, we estimate the electron temperature to be $\sim$85 K, and the dimension of the active region was set to be 10 $\mu$m wide and 2 $\mu$m long to make the device resistance to be $\sim$ 50 $\Omega$ at the bias point based on the sheet resistance of $\sim$250 $\Omega/\Box$ measured with a Hall bar sample at 85 K. The IV curve for the active region was then scaled from the measured IV curve of the Hall bar sample assuming a constant current density (see Figure \ref{fig:sfig7to8}(a)). For the calculation, conventional bolometric mixer theory for semiconducting and superconducing hot-electron bolometric mixers\cite{Arams1966,Karasik,Karasik1996} was used assuming a perfectly matched IF chain, optimal LO power, and a RF coupling efficiency of 70\%, which is a typical value for the optical coupling efficiency in HEB mixers coupled with planar antenna structures.\cite{BorisCoupling}

The expression for the conversion efficiency of a bolometric mixer at the angular IF frequency $\omega_{IF}$,  $\eta(\omega_{IF})$, is given in Ref. \onlinecite{Karasik}: 
\begin{equation}\label{conversioneff}
    \eta\left(\omega\right) = 2\gamma C^{2} \frac{RR_{L}}{\left[Z(\infty)+R_{L}\right]^{2}}\frac{1}{\left[1+C\frac{R-R_{L}}{Z(\infty)+R_{L}}\right]^{2}}\frac{1}{1+\left(\omega_{IF}\tau^{*}\right)^{2}}
\end{equation}
where $\gamma$ is the RF coupling efficiency, $C=\frac{Z-R}{Z+R}$ is the self-heating parameter where $Z$ is the differential resistance and $R$ is the linear resistance in the device IV curve, $R_{L}$ is the load resistance for the IF chain, $Z(\infty)$ is the differential resistance at high IF frequency ($\omega_{IF}\rightarrow\infty$) which we assume $Z\left(\infty\right)=R$, and $\tau^{*}$ is the apparent time constant taking into account the electro-thermal feedback on the thermal relaxation time constant $\tau_{e}$ given by $\tau_{e} = C_{e}/G$ where $C_{e}$ is the heat capacity of the 2DEG and $G$ is the thermal conductivity. Under the assumption of a perfectly matched IF chain ($R=R_{L}$) with $Z(\infty)=R$ and $\omega_{IF}\tau^{*}\ll 1$, the expression given in Eq. \ref{conversioneff} is simplified to
\begin{equation}
    \eta(\omega) \approx \frac{1}{2}\gamma C^{2} = \frac{1}{2}\gamma \left(\frac{Z-R}{Z+R}\right)^{2}
\end{equation}
where we have $\gamma = 0.7$, and $Z$ and $R$ can be determined from the model IV curve. The conversion efficiency was extracted from the IV curve shown in Figure \ref{fig:sfig7to8}(a) and is plotted in terms of the conversion loss $L=10\log \frac{1}{\eta\left(\omega\right)}$ in Figure \ref{fig:sfig7to8}(b). At the source-drain bias of $\sim$ 2.5 mV, the conversion loss is estimated to be $\sim$7.5 dB. 

For the single-sideband (SSB) noise temperature, we consider the contributions from the Johnson noise and the thermal energy fluctuations. The expression for the SSB Johnson mixer noise temperature at $\omega_{IF}$,  $T_{m}^{J}(\omega_{IF})$, and the thermal energy fluctuation mixer noise temperature, $T_{m}^{TF}(\omega_{IF})$, are given in Ref. \onlinecite{Karasik}: 
\begin{equation}
    T_{m}^{J}\left(\omega_{IF}\right) = \frac{2 T_{e} P_{DC}}{\gamma^{2} C^{2} P_{LO}}\frac{ Z\left( \infty \right)}{R} \left[ 1 + \left( \omega_{IF} \tau_{e} \right)^{2} \right]
\end{equation}
and 
\begin{equation}
    T_{m}^{TF}\left(\omega_{IF}\right) = \frac{T_{e}^{2} G}{\gamma^{2} P_{LO}}
\end{equation}
where $T_{e}$ is the electron temperature and $P_{LO}$ is the incident LO power. At $P_{DC}=\gamma P_{LO}$ and with $Z\left(\infty\right)=R$, $\omega_{IF}\tau_{e} \ll 1$ and $G=\alpha P_{DC}/C$ where $\alpha$ is the temperature coefficient of resistance, these expressions are simplified to
\begin{equation}
    T_{m}^{J}\left(\omega_{IF}\right) \approx \frac{2 T_{e}}{\gamma}\left(\frac{Z+R}{Z-R}\right)^{2}
\end{equation}
and 
\begin{equation}
    T_{m}^{TF}\left(\omega_{IF}\right) \approx \frac{T_{e}^{2}\alpha}{\gamma}\frac{Z+R}{Z-R}
\end{equation}
where $T_{e}$ is estimated to be 85 K (at the lattice temperature $T$ = 60 K), $\alpha$ is measured to be $\sim$0.02 K\textsuperscript{-1}, and $\gamma$ is 0.7. The total SSB mixer noise temperature $T_{m}=T_{m}^{J}+T_{m}^{TF}$ is plotted in Figure \ref{fig:sfig7to8}(b). At the bias point of $\sim$ 2.5 mV at which the dissipated DC power is $\sim$0.15 $\mu$W, the SSB mixer noise temperature is estimated to be $\sim$700 K, and the LO power is $\sim$0.2$\mu$W.

\bibliography{TACIT}

\end{document}